\pdfoutput=1
\documentclass[reprint,aps,prd,amsmath,showpacs]{revtex4-1}
\usepackage{graphicx}
\usepackage[pdftex,linkcolor=blue,citecolor=blue]{hyperref}
\usepackage{bm}
\usepackage{slashed}
\usepackage{float}
\usepackage[caption=false]{subfig}
\usepackage{amssymb}
\usepackage{latexsym}
\usepackage{natbib}
\hypersetup{colorlinks=true}

\newcommand{\de}{\ensuremath{\mathrm{d}}}
\newcommand{\elel}{\ensuremath{\ell^+\ell^-}}

\newcommand{\eeg}{\ensuremath{e^+e^-\gamma}}
\newcommand{\ee}{\ensuremath{e^+e^-}}
\newcommand{\tta}{\ensuremath{\tau^+\tau^-}}
\newcommand{\mm}{\ensuremath{\mu^+\mu^-}}

\newcommand{\kp}{\ensuremath{k_{\perp}}}
\newcommand{\vkp}{\ensuremath{\bm{k}_{\perp}}}

\begin{document}
\title{High Resolution Nonperturbative Light-Front Simulations of the
True Muonium Atom}

\author{Henry Lamm}
\email{hlammiv@asu.edu}
\author{Richard F. Lebed}
\email{Richard.Lebed@asu.edu}
\affiliation{Department of Physics, Arizona State University, Tempe,
AZ 85287, USA}
\date{\today}

\begin{abstract}
  Through the development of a parallel code called TMSWIFT, an
  extensive light-front quantization study of the nonperturbative
  spectrum of the bound state $(\mm)$, true muonium, has been
  performed.  Using Pad\'{e} approximants, it has been possible to
  extract continuum and infinite-cutoff limits for the singlet and
  triplet states for a range of values of the coupling constant
  $\alpha$.  This data set allows for an investigation of the $\alpha$
  dependence of the light-front spectra, the results of which are
  compared to standard calculations.  Decay constants have also been
  obtained.  Improved calculations have been undertaken for the energy
  shifts due to the presence of a second, lighter flavor ($e$).
  Finally, initial results for three-flavor ($e$, $\mu$, $\tau$)
  calculations are presented.
\end{abstract}

\maketitle
\section{Introduction}
\label{sec:intro}
True muonium is the as-yet undiscovered $(\mm)$ bound state.  Its
spectrum, with lifetimes in the range of ps to
ns~\cite{Brodsky:2009gx}, is well defined since the 2.2 $\mu$s
weak-decay lifetime of the muon is much longer.  The levels and
transitions of true muonium are dominated by QED effects because its
purely leptonic nature relegates the influence of QCD to vacuum
polarization, where it contributes a small effect at
$\mathcal{O}(\alpha^5)$~\cite{Jentschura:1997tv,Jentschura:1997ma}.
Electroweak effects are suppressed further, and become relevant only
when all $\mathcal{O}(\alpha^7)$ terms are
considered~\cite{PhysRevD.91.073008}.  The existing discrepancies in
muon physics (the muon anomalous magnetic moment
$(g-2)_\mu$~\cite{PhysRevD.73.072003}, the proton charge radius
$r_P$~\cite{Antognini:1900ns}, $B^+\rightarrow K^+\elel$
decays~\cite{Aaij:2014ora}, $\bar{B}^0 \to
D^{*+}\ell^{-}\bar{\nu}_{\ell}$~\cite{Aaij:2015yra}) motivate a
serious investigation of true muonium, which has been shown to have
strong discriminating power among alternative resolutions to these
anomalies~\cite{TuckerSmith:2010ra,
  PhysRevD.91.073008,Lamm:2015fia,Lamm:2015gka}.  Using the methods
developed in this paper, nonperturbative corrections to bound states
from these new physics proposals could be investigated through the
inclusion of new matrix elements, allowing for more stringent
constraints than those obtained through conventional perturbative
studies.

The atom's non-observation to date is due to difficulties in producing
associated low-energy muon pairs, as well as its short lifetime.  Many
proposed methods of production exist~\cite{Nemenov:1972ph,
Moffat:1975uw,Holvik:1986ty,Kozlov:1987ey,Ginzburg:1998df,
ArteagaRomero:2000yh,Brodsky:2009gx,Chen:2012ci,Banburski:2012tk,
Ellis:2015eea}.  The Heavy Photon Search (HPS) experiment in 2016 will
begin a search for true muonium at a fixed
target~\cite{Celentano:2014wya,Banburski:2012tk}.  Additionally, the
DImeson Relativistic Atom Complex (DIRAC) might observe the atom in an
upgraded run~\cite{Benelli:2012bw,dirac}.  Given enough statistics,
DIRAC could obtain a value for the Lamb shift using methods developed
for ($\pi^+\pi^-$)~\cite{Nemenov:2001vp}.  These experiments produce
relativistic true muonium: In general, the $\mu^+$ and $\mu^-$ are
produced relativistically, both with respect to the lab frame and each
other.  Unfortunately, instant-form (conventional fixed-time) wave
functions are not functions of boost-invariant variables (because the
$\mu^+$ and $\mu^-$ rest frames are not the same); thus, production
and decay rates can be modified.  To reduce this uncertainty, we
produce boost-invariant wave functions through light-front
techniques~\cite{Lamm:2013oga,Lamm:2016ong}.

To establish the context for the work presented here, a discussion of
the history of this and related problems is appropriate.  Weinberg,
interested in the \textit{infinite-momentum frame} (in which a state's
momentum component $p_z\rightarrow\infty$), discovered in the case of
the $\phi^3$ theory that creating or annihilating particles from the
vacuum was forbidden~\cite{Weinberg:1966jm}.  This observation
eventually led to the understanding that the vacuum of such a field
theory is trivial ({\it i.e.}, empty of ordinary particles), and that
Fock states with fixed particle content are well-defined.  Instead of
taking the infinite-momentum limit of instant-form field theory, one
can obtain equivalent results by quantizing at fixed values of
light-front time $x^+ \equiv t + z$ (called {\it front
  form})~\cite{Dirac:1949cp}.  In front form, one is able to develop a
rigorous, closed-form Hamiltonian formalism~\cite{Brodsky:1997de}.  In
this formalism, an analogue of the Schr\"{o}dinger equation exists,
since an infinite but denumerable set of coupled integral equations
for eigenstates of the Hamiltonian occurs.  The front form admits a
perturbation theory, and its Feynman rules were derived by Kogut and
Soper~\cite{Kogut:1969xa}.  Because of the inequivalent nature of
instant-form and front-form quantization, it has been a crucial, but
highly nontrivial, matter to show that the traditional instant-form
calculations give equivalent results to those from the front
form~\cite{Ligterink:1994tm,
  Chang:1972xt,Yan:1973qf,Chang:1973qi,Bakker:2002ih,Bakker:2005mb,
  Misra:2005dt,Bakker:2006pn,Patel:2010vd}.

Perturbative front-form methods have shown success in the study of
non-Abelian gauge theories.  In QCD, these methods have been used to
obtain results for exclusive processes by Lepage and
Brodsky~\cite{Lepage:1980fj}, where equivalent instant-form
expressions did not exist.  Analytical results using light-front
techniques have also reproduced the correct leading-order Lamb shift
and hyperfine splitting for QED bound
states~\cite{Jones:1996pz,Jones:1996vy, Jones:1997cb,Jones:1997iu}.
The Yukawa theory has been used to understand the differences between
instant-form and front-form approaches and how they can be
reconciled~\cite{Glazek:1992bs,
  Bakker:2006pn,Bakker:2002ih,ManginBrinet:2001tc,ManginBrinet:2003nm}.
The complete Standard model has also been formulated in light-front
quantization~\cite{PhysRevD.66.045019}.

As mentioned, the existence of a closed-form Hamiltonian allows for a
Schr\"odinger-like equation that can be expressed in an
infinite-dimensional Fock space, which can be used to solve
nonperturbative field theory, and allows techniques from
nonrelativistic quantum mechanics to be applied to quantum field
theory.  To make these problems tractable, the infinite set of coupled
equations must be truncated in a suitable way.  In analogy to results
in instant form, these truncations can produce divergent results and
must be regularized to obtain sensible answers.  The topic of how to
renormalize such a Hamiltonian was first considered
in~\cite{Glazek:1993rc}.  One method proceeds by truncating the Fock
space to a finite number of states based on particle content.  In this
truncation, renormalization is possible through Fock state
sector-dependent counterterms~\cite{Karmanov:2001te,
  ManginBrinet:2001tc, Carbonell:2002xs,ManginBrinet:2003nm,
  Karmanov:2008br,Karmanov:2010ih, Karmanov:2010zz,Karmanov:2012aj},
Pauli-Villars regulators~\cite{Chabysheva:2009vm,Chabysheva:2010vk,
  Malyshev:2013eca,Hiller:2015bic,Chabysheva:2015vga}, or the use of
flow equations~\cite{Gubankova:1998wj,Gubankova1998,Gubankova:1999cx,
  Gubankova:2000cia,Gubankova:2000ci}.  While each method works in
principle, the practical difficulty of renormalizing nonperturbative
Hamiltonians remains daunting.

In order to solve these field theories numerically, the Fock states
are furthermore discretized in momentum Fourier modes on a lattice, a
method called Discretized Light Cone Quantization (DLCQ).  This method
was pioneered by Pauli and Brodsky, working with a 1+1-dimensional
Yukawa theory~\cite{Pauli:1985ps}.  The special feature of
super-renormalizability of field theories in 1+1 has been particularly
amenable to DLCQ, and these theories have been investigated in depth.
Sawicki used the method to solve scalar $\rm
QED_{1+1}$~\cite{Sawicki:1985uq,Sawicki:1985vs}, while Harindranath
and Vary investigated the structure of the vacuum and bound states of
$\phi^3_{1+1}$ and $\phi^4_{1+1}$ models~\cite{Harindranath:1987db,
  Harindranath:1988zt,Harindranath:1988zs}.  Pushing further,
Hornbostel {\it et al.\/} presented results for the meson and baryon
eigenstates of $\rm QCD_{1+1}$~\cite{Hornbostel:1988fb}, while Swenson
and Hiller studied more field-theoretical properties of the
light-front in the Wick-Cutkosky model~\cite{Swenson:1993ci}.  The
Schwinger model, which admits analytical solutions in both instant
form and front form, was first studied by Eller {\it et al.\/} in
1986~\cite{Eller:1986nt}, and since has become an important test bed
for developing improvements that can then be used in other
theories~\cite{Heinzl:1991dw,
  McCartor:1991gn,McCartor:1994im,McCartor:1996pe,Nakawaki:1999ee,
  Strauss:2008zx,Strauss:2009zza,Strauss:2010zza}.

Since DLCQ produces both the wave functions and the energy levels,
Hiller was able to compute the $R$-ratio in $\rm
QED_{1+1}$~\cite{Hiller:1990xy}.  In one spatial dimension, DLCQ has
also been applied to solving 't~Hooft's model of large-N
QCD~\cite{vandeSande:1996mc}, adjoint QCD~\cite{Trittmann:2000uj,
  Trittmann:2001dk,Trittmann:2015oka}, and supersymmetric
models~\cite{Matsumura:1995kw,Antonuccio:1998kz,Antonuccio:1998jg,Antonuccio:1998tm,
Antonuccio:1998mq,Antonuccio:1998zp,Antonuccio:1998mw,Antonuccio:1998zu,
Antonuccio:1999jj,Haney:1999tk,Lunin:1999ib,Antonuccio:1999iz,Lunin:2000im,
Pinsky:2000rn,Hiller:2000nf,Hiller:2001mh,Hiller:2001tv,Hiller:2002cg,
Hiller:2002pj,Hiller:2003qe,Hiller:2003jd,Hiller:2004rb,Hiller:2005vf,
Hiller:2007sc,Trittmann:2009dw}.  Although spontaneous symmetry
breaking is manifested in a distinctly different way in 1+1, it is
also possible to study using DLCQ~\cite{Pinsky:1993yi,Pinsky:1994si,
  Bender:1992yd}.  Finally, research has been undertaken using DLCQ to
test Maldecena's AdS/CFT conjecture in 1+1
theories~\cite{Antonuccio:1999iz,Hiller:2000nf, Hiller:2005vf}.

Extending DLCQ beyond 1+1 dimensions is complicated in two ways:
first, higher-dimensional theories require regularization and
renormalization, as discussed above.  Second, the number of Fock
states grows so rapidly that tractable numerical calculations allow
only a small number of states to be included.  Despite these
difficulties, DLCQ was applied first to positronium by Tang {\it et
  al.}~\cite{Tang:1991rc}.  In that work, the effective Hamiltonian
matrix equation was derived for a model including only the
$|\ee\rangle$ and $|\eeg\rangle$ Fock states.  Variational methods
were applied to this effective model and produced upper limits on the
triplet state.  Attempts to apply DLCQ to QCD were undertaken at the
same time by Hollenberg~\cite{Hollenberg:1991fk}, but renormalization
and computational resources prevented much success.  Further
developments in understanding the connection between light-front and
instant-form techniques were studied by Kalu\v{z}a and Pauli,
reproducing the expected results for the hyperfine splitting and Bohr
states in the limit of $\alpha\rightarrow0$~\cite{Kaluza:1991kx}.
Krautg\"artner {\it et al.}, implementing the \textit{Coulomb
  counterterm} techniques developed by W\"olz~\cite{Wolz:1990zzz},
solved the effective matrix equation for
positronium~\cite{Krautgartner:1991xz}.  They found that it was
possible to reproduce the correct Bohr spectrum, as well as the
leading relativistic hyperfine splitting, for both $\alpha_{\rm
  QED}=1/137$ and $\alpha=0.3$, albeit with some cutoff dependence.
Concerned with the effect of \textit{zero modes\/} (nontrivial field
configurations in the Fock vacuum), Kalloniatis and Pauli undertook
numerical simulations based upon perturbative solutions to the
zero-mode constraint equations~\cite{Kalloniatis:1993jh}.

Krautg\"artner further developed these techniques and began to
analytically study the two-photon exchange interaction and its
relationship to the observed divergences in his
dissertation~\cite{Krautgartner:1992zzz}.  W\"olz, in his
dissertation, applied DLCQ to QCD by including the
$|q\bar{q}gg\rangle$ Fock state~\cite{Wolz:1995zzz}.  Numerical
limitations at the time prevented implementation of the counterterm
techniques being concurrently developed, so that a slow convergence in
the number of discretization points and a strong dependence on the
momentum cutoff precluded these results from suggesting any conclusive
statements.  Synthesizing all these techniques, Trittmann computed the
first results for positronium with the inclusion of the annihilation
$\ee\rightarrow\gamma$
channel~\cite{Trittmann:1997xz,Trittmann:1997tt,Trittmann:2000gk}.
Utilizing the good quantum number $J_z$, he was able to split the
problem into sectors and investigate the breaking of rotational
invariance inherent in light-front form in the effective equation.
Cutoff dependence and inadequate computational resources were the
major limits to Trittmann's work.  With improved computing resources
and the introduction of a special counterterm to cancel a divergent
matrix element, DLCQ was applied by the current authors to two-flavor
QED to obtain bound states of positronium and true muonium
simultaneously~\cite{Lamm:2013oga}.  This work built upon the prior
methods by incorporating a number of features of QED bound states in
front-form field theory.

Beyond DLCQ, other numerical methods have been developed for
light-front systems.  Basis light-front quantization (BLFQ) follows
from discretizing the momenta into harmonic-oscillator modes in the
transverse direction instead of using Fourier modes.  This method
aspires to decrease the number of basis states needed by more
accurately representing the functional behavior of the wave function.
BLFQ has shown initial success in solving bound-state problems in
QED~\cite{Vary:2009gt,Zhao:2011ct,Maris:2013qma,Zhao:2013cma,
  Zhao:2013jia,Chakrabarti:2014cwa,Zhao:2014xaa} and
QCD~\cite{Li:2015zda}.  Using Monte Carlo methods developed for
instant-form lattice gauge theory, transverse lattice theory has
investigated simple models of QCD in 3+1
dimensions~\cite{Pauli:1995dt,Dalley:1996ns,vandeSande:1996gh,
  Dalley:1999ay}.  Tube-based, collinear QCD and other
effective-Hamiltonian methods also exist~\cite{vandeSande:1995cg,
  Brisudova:1996vw,Chakrabarti:2001yh}.  In recent years, the AdS/QCD
conjecture has been extended to light-front field theory to produce
the low-energy meson and baryon spectra~\cite{deTeramond:2008ht,
  deTeramond:2010we,Brodsky:2010ur,deTeramond:2010ge,Brodsky:2013ar,
  Brodsky:2014yha,deTeramond:2014asa,Brodsky:2015oia,Dosch:2015bca,
  Dosch:2015nwa}.

The limitations of Fock-state truncation in renormalization have also
prompted the study of other methods of truncation.  Drawing upon the
techniques found in many-body physics,
coupled-cluster~\cite{Chabysheva:2011ed, Chabysheva:2012bf,
Elliott:2014fsa} and coherent-basis truncations~\cite{Misra:1993ps,
Misra:1995sd,Misra:2000ub,Misra:2005dt,More:2012ey,More:2013lua} have
shown promise in simpler systems.

This paper is organized as follows.  In Sec.~\ref{sec:model} we
briefly review the model of true muonium studied here.
Section~\ref{sec:result} is devoted to presenting the numerical
results obtained for the energy levels and the decay constants, with
emphasis on the effect of the annihilation channel and of the presence
of multiple flavors on the states.  We conclude in Sec.~\ref{sec:cons}
with some discussion of our results and possible directions for future
work.

\section{True Muonium model}
\label{sec:model}

We review here the major points of our model, which are described in
detail in a previous work~\cite{Lamm:2013oga}.  In front form, the
eigenvalue equation for a bound state is given by:
\begin{align}
\label{eq:ham}
\bigg(M^2 - \sum_i&\frac{m^2_i+\bm{k}^{2}_{\perp i}}{x_i} \bigg)
\psi (x_i,\bm{k}_{\perp i};h_i)\nonumber\\ =\sum_{h_j}&\int_D
\mathrm{d}x'_{j} \mathrm{d}^2 \bm{k}'_{\perp j}
\langle x_i,\bm{k}_{\perp i};h_i \left| V_{\rm eff}
\right| x'_{j}, \bm{k}'_{\perp j};h_j \rangle\nonumber\\&\times \psi
(x'_{j},\bm{k}'_{\perp,j}; h_j),
\end{align}
where $M$ is the invariant mass of the state, $m$ indicates a mass
term, $i,j$ are component particle indices, $x$ and $\bm{k}_\perp$ are
the conventional longitudinal and transverse momentum light-front
coordinates, respectively, $h$ is shorthand for all intrinsic quantum
numbers of a state, and $V_{\rm eff}$ are interaction terms given by
the light-front Hamiltonian.  The domain $D$ of Eq.~(\ref{eq:ham}) is
made well defined by the introduction of cutoff $\Lambda$, and we
choose~\cite{Lepage:1980fj}
\begin{equation}
\frac{ m^2 + \bm{k}_\perp^2 }{x(1-x)} \le \Lambda^2 + 4m^2 \, .
\end{equation}
Our model considers only the truncated Fock space of
$|\ell_i\bar{\ell}_i\rangle$, $|\ell_i\bar{\ell}_i\gamma\rangle$, and
$|\gamma\rangle$.  The single-photon interaction allows for mixing
between flavors via the annihilation channel.  The wave functions are
in the form of helicity states only for pure lepton states ({\it
e.g.}, $\left|\mm\right>$).  The $\left|\gamma\right>$ and
$\left|\ell_i\bar{\ell}_i\gamma\right>$ components are folded into
$V_{\rm eff}$ by means of the method of iterated
resolvents~\cite{Pauli:1997ns,Trittmann:1997xz}.

Discretization in ($x,\bm{k}_\perp$) space results in an asymmetric
matrix in the discretized form of Eq.~(\ref{eq:ham}), which
significantly increases the computational effort, so instead it is
numerically superior to use the polar coordinates utilized initially
by Karmanov~\cite{Karmanov:1980mc} to study a toy model of the
deuteron, and later by Sawicki~\cite{Sawicki:1985vs,Sawicki:1985uq} in
studying relativistic scalar-field bound states on the light front.
These coordinates are defined by
\begin{equation} \label{eq:xdefn}
 x=\frac{1}{2}\left(1+\frac{\mu\cos \theta }
{\sqrt{m_i^2+\mu^2}}\right) \, ,
\end{equation}
\begin{equation} \label{eq:kperpdefn}
 \bm{k}_\perp = \mu ( \sin \theta \cos \phi ,
\sin \theta \sin \phi , 0 ) \, .
\end{equation}
Using these variables, one may exchange $\phi$ for the discrete
quantum number $J_z$~\cite{Trittmann:1997xz} and compute using only
$\mu$, $\theta$.  The new variable $\mu$ can be considered an
off-shell momentum, due to the relation
\begin{equation} \label{eq:mudef}
 \frac{m_i^2+\bm{k}^2_\perp}{x(1-x)}=4(\mu^2+m_i^2) \, .
\end{equation}
Since these coordinates depend upon the fermion mass $m_i$, different
sets of $\mu$, $\theta$ values result from the same sets of $x$ and
$\bm{k}_\perp$ values in the multiple-flavor system.

It has been shown~\cite{Krautgartner:1991xz,Trittmann:1997xz,
Lamm:2013oga} that strong dependence in ${}^1S_0$ states on $\Lambda$
arises from the matrix element between antiparallel-helicity states
called $G_2$.  In the limit of $k_\perp \equiv |\bm{k}_\perp|$ or
$k'_\perp \equiv |\bm{k}'_\perp|\to
\infty$, this interaction approaches
\begin{equation}
\label{eq:g2lim}
 \lim_{\kp\rightarrow
 \infty}G_2=-\frac{\alpha}{\pi}\frac{2}{x+x'-2xx'}\delta_{J_z,0} \, ,
\end{equation}
which, in the absence of the dependence of $|\psi_{\elel}\rangle$ upon
$\kp$, would result in a $\delta$ function-like behavior in
configuration space.  Reference~\cite{Krautgartner:1991xz} chose to
regularize this singularity by deleting the entire divergent term.
Instead, a numerically superior subtraction scheme is obtained by only
removing its limit as $k_\perp$ or $k'_\perp \to \infty$,
\begin{equation}
\label{eq:greg}
 G_{2,\rm reg}=G_2+\bigg\{\frac{\alpha}{\pi}\frac{2}{x+x'-2xx'}
\delta_{J_z,0}\bigg\},
\end{equation}
which retains part of the term (including $x$ and $x'$ dependence).
This scheme removes the strongest $\Lambda$ dependence of ${}^1S_0$
states in both QED~\cite{Lamm:2013oga,Wiecki:2014ola} and
QCD~\cite{Li:2015zda} models.  It is important to note that the $\vkp$
dependence of $|\psi_{\elel}\rangle$ varies with $\alpha$, and
therefore it should be anticipated that the strength of this apparent
divergence should also depend upon $\alpha$.  With this regularization
scheme, the model allows for taking the $\Lambda\rightarrow\infty$
limit, albeit with a regularization dependence determined by
mathematical, rather than purely physical, considerations.

Much of the previous work on QED with DLCQ has focused upon the
unphysically large value $\alpha = 0.3$.  In this regime, QED
perturbative calculations can potentially become unreliable.  We use
this strong coupling value of $\alpha$ to study flavor mixing.  New to
this work, we investigate the approach to the physical
$\mathcal{O}(10^{-2})$ value of the QED coupling constant.

\section{Results}
\label{sec:result}
Previous work has given results sensitive to numerical artifacts,
limiting the reliability of the results that could be obtained.  To
overcome some of these limitations, we have produced a new numerical
code, TMSWIFT (True Muonium Solver With Front-from Techniques), which
is available online~\cite{TMSWIFT}.  This code uses the parallel
eigenvalue-solver package SLEPc~\cite{Hernandez:2005:SSF}, both to
increase the number of Fock states and to decrease the time of
calculation.  TMSWIFT allows an arbitrary number of flavors, each
specified by a distinct mass $m_i$, cutoff $\Lambda$, and
discretization numbers $N_\mu$ and $N_\theta$ (although throughout
this work we will fix $N_\mu=N_\theta=N$).  Different discretization
schemes are available in TMSWIFT for exploration of numerical errors
and efficiency.  Our code also allows easy implementation of new
effective interactions ({\it e.g.}, from $|\gamma\gamma\rangle$
states).  These improvements have also allowed us to investigate lower
values of $\alpha$, where the extrapolation to
$\Lambda,N\rightarrow\infty$ becomes more difficult.  In order to
examine these limits, except for Subsec.~\ref{subsec:Multiple} which
explicitly studies multiple-flavor effects, we restrict ourselves to
the case of single-flavor true muonium.

In this section, we explore a number of properties of true muonium,
dedicating a subsection to each: the invariant squared mass $M^2_n$,
the ground-state hyperfine splitting, the singlet and triplet wave
functions, the decay constants, and multiflavor effects.

\subsection{Invariant Squared Mass}
With larger $N$ and improved regularization, we found it possible to
fit the energy levels, $M^2_n$, to Pad\'{e} approximants of second
order.  To perform these fits, we first fit the $N$ dependence for
each value of $\Lambda$ for which simulations were computed:
\begin{equation}
 \label{eq:mfit}
 M^2(N,\Lambda)=\frac{M^2({\Lambda})+\frac{b}{N}+\frac{c}{N^2}}
{1+\frac{d}{N}+\frac{e}{N^2}}\, .
\end{equation}
Then, the final $N \to \infty$ and $\Lambda \to \infty$ results can be
obtained from a second fit to:
\begin{equation}
 \label{eq:mfit2}
 M^2({\Lambda})=\frac{M^2_\infty+\frac{f}{\Lambda}+\frac{g}{\Lambda^2}}
{1+\frac{h}{\Lambda}+\frac{i}{\Lambda^2}}\, .
\end{equation}
These functions are well defined separately in the
$N\rightarrow\infty$ and $\Lambda\rightarrow\infty$ limits, and
therefore one can extract the continuum- and cutoff-independent
values, $M^2_\infty$.  While in principle the entire data set could be
simultaneously fit in $N$ and $\Lambda$, the large cancellations that
can occur between Pad\'{e} coefficients, and the large number of
parameters to fit in practice, make the process more difficult, and
initial conditions for the fit must be carefully chosen to avoid local
minima of the fits.  Moreover, the two parameters have different
origins: $N$ is a numerical artifact, while $\Lambda$ is a theoretical
artifact.  By fitting separately, these issues are largely avoided.
Results for the ground-state singlet and triplet states are tabulated
in Table~\ref{tab:esp}.
\begin{table*}[t]
\caption{Extrapolated results for the bound-state invariant
squared mass $M^2$ in units of $m_\mu^2$, and the decay constants
$f_V$, $f_P$ in units of $m_\mu$, for a range of $\alpha$ values.  The
column labeled $C_{\rm HFS,LF}$ is the computed hyperfine coefficient
$C_{\rm HFS}$ from Eq.~(\ref{eq:ffchf}).  The column labelled $C_{\rm
HFS,ET}$ is the instant-form prediction for $C_{\rm HFS}$ from
Eq.~(\ref{eq:ifchf}).}
\label{tab:esp}
\begin{center}
\begin{tabular}{l|c|c|c|c|c|c}
\hline\hline
$\alpha$&$M^2(1^1S_0)$&$f_V(1^1S_0)$&$M^2(1^3S_1)$&$f_P(1^3S_1)$&$C_{\rm HFS,LF}$&$C_{\rm HFS,ET}$\\ \hline
0.01 & 3.99989993(3) & $4.18(10)\times10^{-5}$ & 3.99989996(3) & $3.893(6)\times10^{-5}$ &0.76(77) &  0.5834\\
0.02 & 3.9995997(2) & $1.1(4)\times10^{-4}$ & 3.9996002(2) & $1.088(7)\times10^{-4}$ &0.79(42) &  0.5837\\
0.03 & 3.9990987(4) & $2.05(9)\times10^{-4}$ & 3.999101(2) & $1.93(6)\times10^{-4}$ &0.74(34) &  0.5841\\
0.04 & 3.998397(4) & $3.15(5)\times10^{-4}$ & 3.998404(5) & $3.07(7)\times10^{-4}$ &0.76(56) &  0.5847\\
0.05 & 3.9974914(4) & $4.466(2)\times10^{-4}$ & 3.9975098(3) & $3.95(2)\times10^{-4}$ &0.74(2) &  0.5855\\
0.07 & 3.995068(3) & $7.404(7)\times10^{-4}$ & 3.9951351(8) & $5.908(5)\times10^{-4}$ &0.7(4) &  0.5877\\
0.1 & 3.98987(6) & $1.273(2)\times10^{-3}$ & 3.990137(3) & $9.16(3)\times10^{-4}$ &0.67(2) &  0.5922\\
0.2 & 3.9576(6) & $3.9(2)\times10^{-3}$ & 3.9614(5) & $1.9(2)\times10^{-3}$ &0.6(2) &  0.6204\\
0.3 & 3.8996(6) & $1.02(3)\times10^{-2}$ & 3.91538(4) & $2.39(2)\times10^{-3}$ &0.49(2) &  0.6735\\
\hline
\end{tabular}
\end{center}
\end{table*}
\begin{figure}
\includegraphics[angle=-90,width=\linewidth]{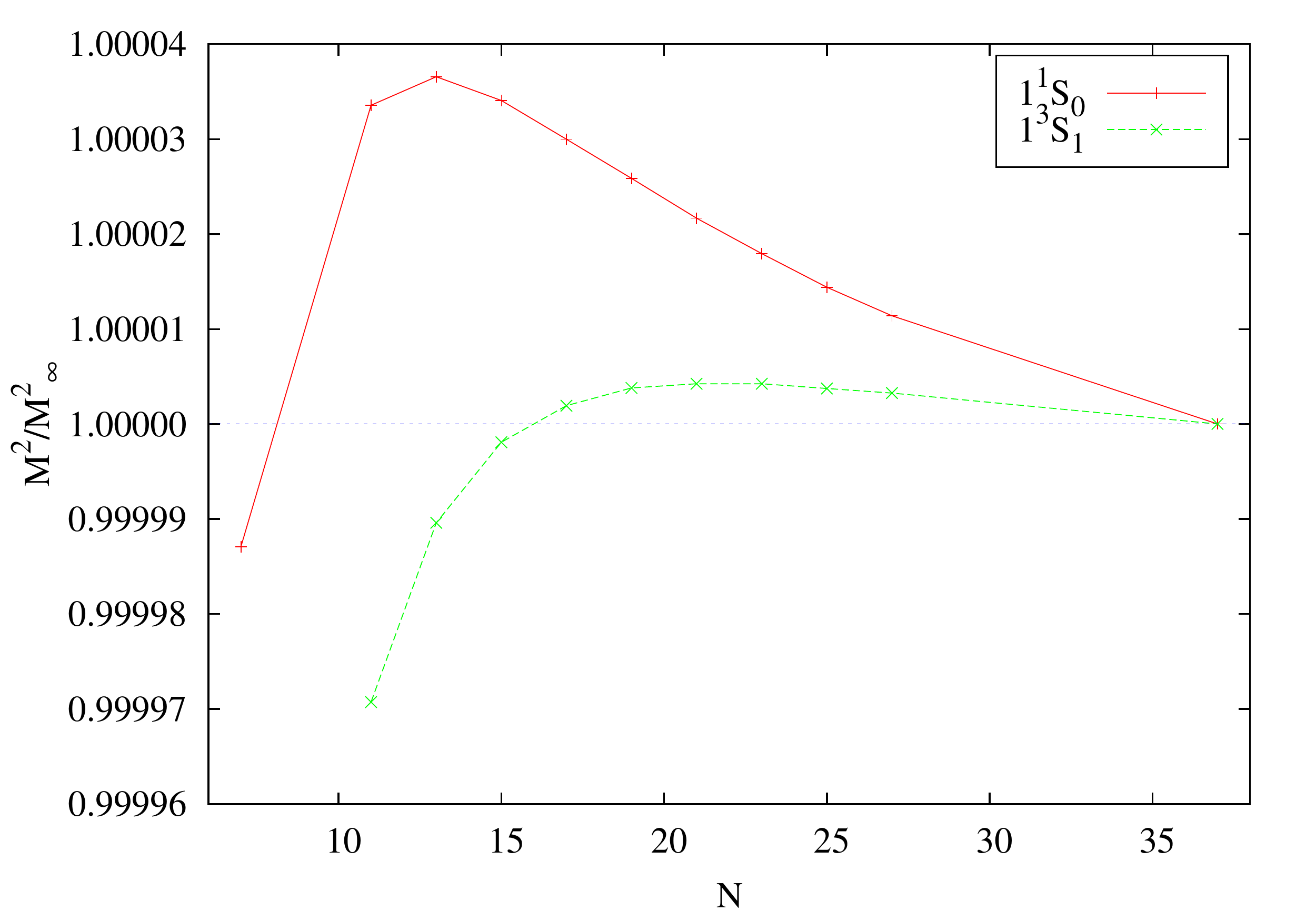}
\caption{\label{fig:ndep}Example of the dependence of $M^2$ upon $N$,
  normalized to the continuum and infinite limit for $\alpha=0.2$,
  $\Lambda = 5m_\mu \alpha$.}
\end{figure}

An example of the dependence of $M^2$ upon $N$ is shown in
Fig.~\ref{fig:ndep} for a fixed value of $\Lambda$ and $\alpha$.  This
dependence is qualitatively the same for all values of $\alpha$ and
$\Lambda$.  As can be seen, with increasing $N$, $M^2$ at first rises
to a peak and then decreases toward a continuum value.  The location
of this peak is found to be proportional to $1/\alpha^2$ and linear in
$\Lambda$.  It is therefore more difficult to numerically simulate
small $\alpha$ and large $\Lambda$, because any results that only
sample to the left of the peaks systematically overestimate $M^2$, by
not respecting that the functions decrease to the right of the peaks.
From Fig.~\ref{fig:ndep} it can also be seen that the triplet state
reaches its (smaller) maximum at a larger $N$.  It is also empirically
found that, while the singlet state peaks at lower $N$, the variance
of $M^2$ is much larger.  With this understanding of the space of $N$,
$\alpha$, and $\Lambda$, one can study the approach to the
perturbative regime of $\alpha$.  The analytic, instant-form values of
$M$ to $\mathcal{O}(\alpha^4)$ are given for $1^1S_0$ and $1^3S_1$
by~\cite{bethe1957quantum}:
\begin{equation}
\label{eq:m2s}
 M^2_{1^1S_0}=\left(2m-\frac{1}{4}m\alpha^2
-\frac{21}{64}m\alpha^4\right)^2,
 \end{equation}
 \begin{equation}
 \label{eq:m2s2}
 M^2_{1^3S_1}=\left(2m-\frac{1}{4}m\alpha^2
+\frac{49}{192}m\alpha^4\right)^2.
\end{equation}
Since $m=1$ in our units, to test these formulas, one can fit to
\begin{equation}
\label{eq:efit}
  M^2(\alpha)=(N_0+N_2\alpha^2+N_4\alpha^4+N_4\alpha^5)^2.
\end{equation}
From the Fock space considered in our model, a perturbative
calculation should not have any higher-order contributions, but one
could anticipate a possible $\mathcal{O}(\alpha^5)$ term due to the
contributions of higher-order terms arising from our nonperturbative
procedure and regularization scheme.  The results of the fit are found
in Table~\ref{tab:Ealpha}.
\begin{table*}[t]
\caption{Parameters of Eq.~(\ref{eq:efit}) for the singlet and triplet
states of true muonium, fit over two ranges of $\alpha$.  The
$\mathcal{O}(\alpha^4)$ perturbative predictions are $N_0=2$,
$N_2=-\frac{1}{4}$, $N_{4,1^1S_0}=-\frac{21}{64}\approx-0.328$,
$N_{4,1^3S_1}=\frac{49}{192}\approx0.255$.  The expected value of
$N_5$ is unknown, but anticipated to be small.  Reported uncertainties
result solely from the fitting procedure.}
\label{tab:Ealpha}
\begin{tabular}{l|l|c|c|c|c}
\hline\hline
$E_n$&$\alpha$&$N_0$&$N_2$&$N_4$&$N_5$\\ \hline
$1^1S_0$ & [0.01,0.3] & 1.99999998(2) &  -0.2500(2) &  -0.37(5) &  -0.04(21)\\
     & [0.01,0.1] & 1.999999990(2) &  -0.25004(2) &  -0.35(2) &  0.08(10)\\
$1^3S_1$ & [0.01,0.3] & 1.99999998(2) &  -0.24990(8) &  0.39(3) &  -0.78(8)\\
     & [0.01,0.1] & 1.999999979(6) &  -0.24993(5) &  0.38(3) &  -0.60(26)\\
\hline
\end{tabular}
\end{table*}

Comparing the singlet-state results to Eqs.~(\ref{eq:m2s}), one sees
that TMSWIFT reproduces within uncertainty the $\mathcal{O}(\alpha^4)$
calculation over the entire range of $\alpha$.  Extracting possible
higher-order coefficients would be possible by increasing $N$ beyond
what has been presented here.  In contrast, for the triplet state,
only the terms up to $\alpha^2$ of Eqs.~(\ref{eq:m2s2}) are correctly
reproduced.  The $\alpha^4$ coefficient reproduced the anticipated
sign, but it is larger than the result of the instant-form
calculation.  Additionally, there is a large, unanticipated $\alpha^5$
coefficient.  Such results are indicative of issues in the
annihilation channel, which affects only the triplet at this order.

\subsection{Hyperfine Splitting}
To study these effects further, one can check how accurately our
front-form model reproduces the expected instant-form results through
the hyperfine coefficient, which is defined as
\begin{equation}
\label{eq:ffchf}
 C_{\rm HFS}\equiv\frac{E_{\rm{HFS}}}{m_\mu\alpha^4}=
\frac{\sqrt{M^2(1^3S_1)}-\sqrt{M^2(1^1S_0)}}{m_\mu\alpha^4}.
\end{equation}
If all Fock states were included in our model, then the full known
$\mathcal{O}(\alpha^7)$ instant-form prediction of $E_{\rm HFS}$ of
Ref.~\cite{PhysRevD.91.073008} could be compared to our results.  But
because of our Fock-state truncations, there is a mismatch in the
higher-order contributions.  Since we can only extract up to
$\mathcal{O}(\alpha^4)$, it is useful to compare to the leading-order
value of $C_{\rm HFS}=\frac{7}{12}$.

Our model would be expected to partially resum the relativistic
corrections from the single-photon exchange and annihilation diagrams.
Therefore, we present the values of $C_{\rm HFS}$ given by the exact
Dirac-Coulomb solutions~\cite{Breit:1930}:
\begin{align}
\label{eq:ifchf}
 C_{\rm HFS}&=\frac{1}{m_\mu\alpha^4}\left(\frac{E_F}
{\sqrt{1-\alpha^2}[2\sqrt{1-\alpha^2}-1]}\right)\nonumber\\
&=\frac{7}{12}\left(1+\frac{3}{2}\alpha^2+\frac{17}{8}\alpha^4
+\mathcal{O}(\alpha^6)\right),
\end{align}
where $E_F=\frac{7}{12}m_\mu\alpha^4$ is the lowest-order hyperfine
splitting of true muonium.  If higher precision could be attained,
these effects might be resolvable, but at the current levels they are
not yet visible.

Previous results for $C_{\rm HFS}$ at $\alpha=0.3$ without the
regularization term are found in Table 4.2 of
Ref.~\cite{Trittmann:1997xz}, and can be calculated from the results
found in Ref.~\cite{Gubankova:1999cx}.  The $C_{\rm HFS}$ obtained in
these works appears to have a logarithmic singularity in the singlet
state, indicating that no $\Lambda\rightarrow\infty$ limit could be
taken.  The severity of the divergence can be seen in
Ref.~\cite{Trittmann:1997xz}, where $C_{\rm HFS}$ rises from
$\approx0.313$ at $\Lambda=m_f$ to $\approx 1.27$ at $\Lambda=18m_f$.
In contrast, we find that for our regularization scheme, $C_{\rm HFS}$
is finite because the two energy levels are finite in the
$N\rightarrow\infty$ and $\Lambda\rightarrow\infty$ limits.
The numerical results in Table~\ref{tab:esp} are roughly consistent
over the entire range of $\alpha$, albeit with large uncertainty.
While the results are finite, we find that the central values are
systematically larger than the anticipated $\frac{7}{12}\approx0.58$,
being in the range $0.7$--$0.8$ except for $\alpha=0.3$, where
observables approach their asymptotic values more slowly due to
changes in the wave function large-$\bm{k}$ dependence, as
discussed in the next section.

Clearly, a disagreement is seen between the two instant-form
predictions and the results on the light front.  Previously, several
authors~\cite{Tang:1991rc,Krautgartner:1991xz,Trittmann:1997xz} have
also pointed out that the correct value of $C_{\rm HFS}$ is best
obtained for $\Lambda\approx m\alpha$, and the results from TMSWIFT
support this point of view.  Unfortunately, the divergences spoil this
agreement at larger $\Lambda$, necessitating renormalization.  The
larger splitting in the infinite-$\Lambda$ limit can be understood
thusly: Although the regularization procedure developed allows for
extrapolation to $\Lambda\rightarrow\infty$, the $\Lambda$ dependences
of the singlet and triplet states are different, as was seen
in~\cite{Lamm:2013oga}, leading to an asymptotic HFS that, while
finite, is larger than the known result.
 
These results are in contrast to the situation in which the
annihilation-channel interaction is excluded.  Choosing the
intermediate case of $\alpha=0.1$, we performed an exploratory search
with a smaller number of simulations.  In this case, we found in the
continuum and infinite-$\Lambda$ limits that $C_{\rm HFS}=0.35(11)$,
in agreement with the anticipated value at leading order of $C_{\rm
  HFS}\approx0.333$.  A similar small study for $\alpha=0.3$ with
$J_z=1$ also found a value of $C_{\rm HFS}\approx0.75$, indicating
that both the dynamical and instantaneous annihilation interactions
are affected.  This evidence further suggests that the
annihilation-channel interaction is the source of the discrepancies.

To understand why the annihilation channel gives trouble, it is useful
to recall how this term is included in instant form.  In standard,
perturbative nonrelativistic calculations, these contributions in
coordinate space are represented as a contact term $\propto
\delta^{(3)}({\bf r})$; therefore, in momentum space these terms are
very sensitive to large momenta, and imposing a cutoff $\Lambda$
prevents these momenta from contributing.  Furthermore, we have
already seen that obtaining numerical results for large $\Lambda$ is
complicated by the need to include much larger $N$ than is currently
possible.  Put together, these facts indicate that regularization and
renormalization is a more complicated affair in the annihilation
channel.

\subsection{Wave Functions}

In order to understand the effect of the regularization term on the
effective interaction, we have studied the large-$\mu$ behavior of the
wave functions.  The momentum-space wave function obtained from the
nonrelativistic Schr\"odinger equation is
\begin{equation}
 \Psi(\bm{k})=\frac{\sqrt{8}}{\pi}\frac{1}{(1+\bm{k}^2)^2},
\end{equation}
where the instant-form 3-momentum carries units of Bohr momentum
$\frac{1}{2}m\alpha$.  It is known that higher-order corrections to
the interaction lead to a modified power law, changing the
large-$\bm{k}$ power scaling from the nonrelativistic value of $-4$.
Since according to Eq.~(\ref{eq:mudef}) $\bm{k}$ is linear in $\mu$,
for our studies it suffices to compute the dependence upon $\alpha$ at
large $\mu$.  The large-$\mu$ behavior is parametrized as
\begin{equation}
\label{eq:kfit}
 \Psi(\mu)=a\mu^{-\kappa},
\end{equation}
where $\kappa=4$ is the result for the nonrelativistic Schr\"odinger
equation.

\begin{figure}
\includegraphics[width=\linewidth]{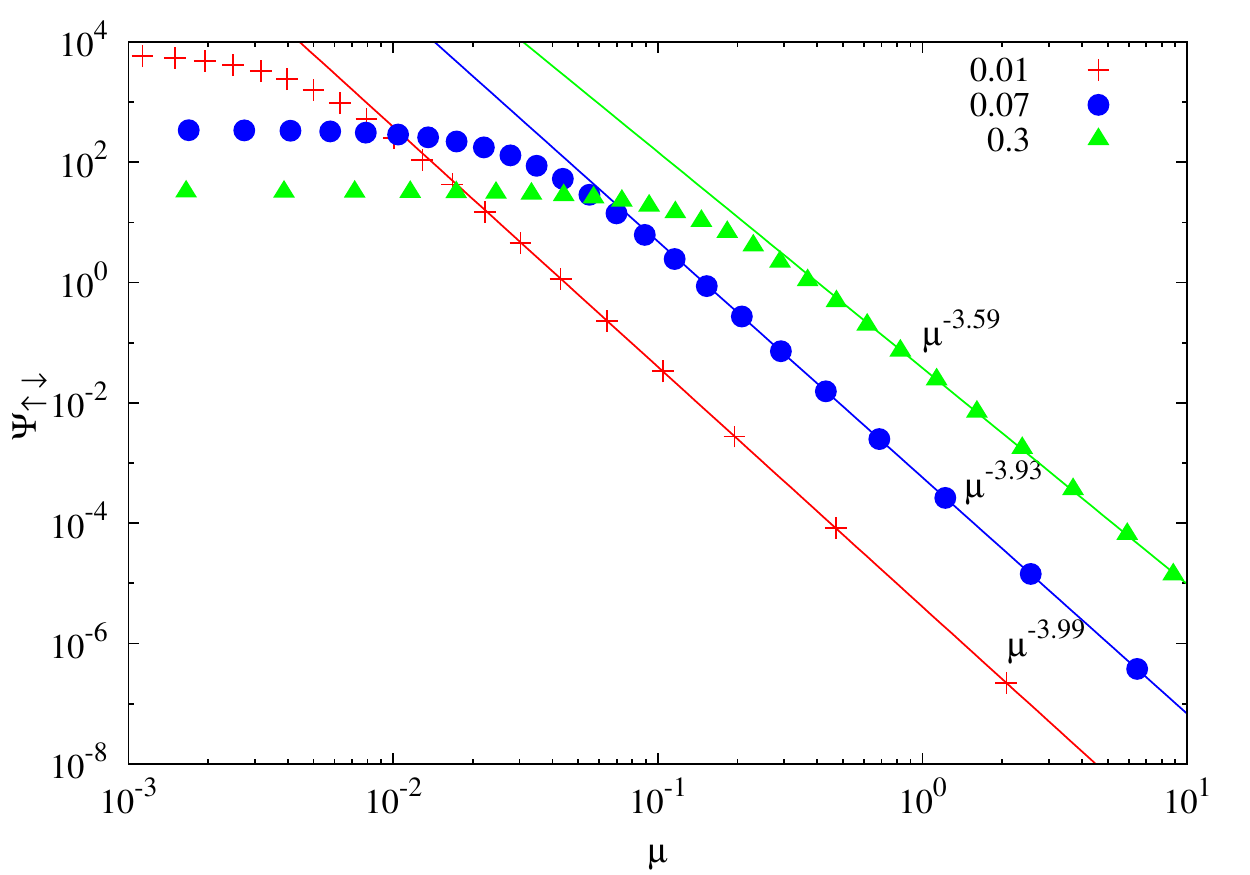}
 \caption{\label{fig:sing}Dependence of the $\uparrow\downarrow$
 component of the $1^1S_0$ state upon $\mu$ for a fixed value of
 $x=0.5$ for different values of $\alpha$.  The points indicate the
 numerical results, and the solid lines are the fits used to extract
 $\kappa$.}
\end{figure}

In~\cite{Krautgartner:1991xz}, it was found that for $\alpha_{\rm
  QED}=1/137$, the large-$\mu$ behavior of the $\uparrow\downarrow$
singlet wave function is $\kappa=4.0$, in agreement with expectations,
and that for $\alpha=0.3$ the behavior is $\kappa=2.5$.  We believe
this large reduction in $\kappa$ is related to the strong $\Lambda$
dependence found in~\cite{Krautgartner:1991xz,Trittmann:1997xz}.  To
further understand the relation between regularization and $\kappa$,
we fit the large-$\mu$ tail of our wave functions to
Eq.~(\ref{eq:kfit}) with the results for a selected few values of
$\alpha$ shown in Fig.~\ref{fig:sing}.  In all of these cases, we have
implemented our regularization subtraction scheme.
The values of $\kappa$ for $\alpha=0.01,0.07$ appear to show only
small deviations from the nonrelativistic value, consistent
with~\cite{Krautgartner:1991xz}.  In contrast, our value of
$\kappa=3.59$ for $\alpha=0.3$ is dramatically larger than found in
the unregulated results of~\cite{Krautgartner:1991xz}.  Since the
large-$\mu$ tail decays much faster than
in~\cite{Krautgartner:1991xz}, the contribution of any potentially
divergent terms will be reduced, explaining why the results
of~\cite{Lamm:2013oga} showed such a dramatic improvement.

\begin{figure}
\includegraphics[width=\linewidth]{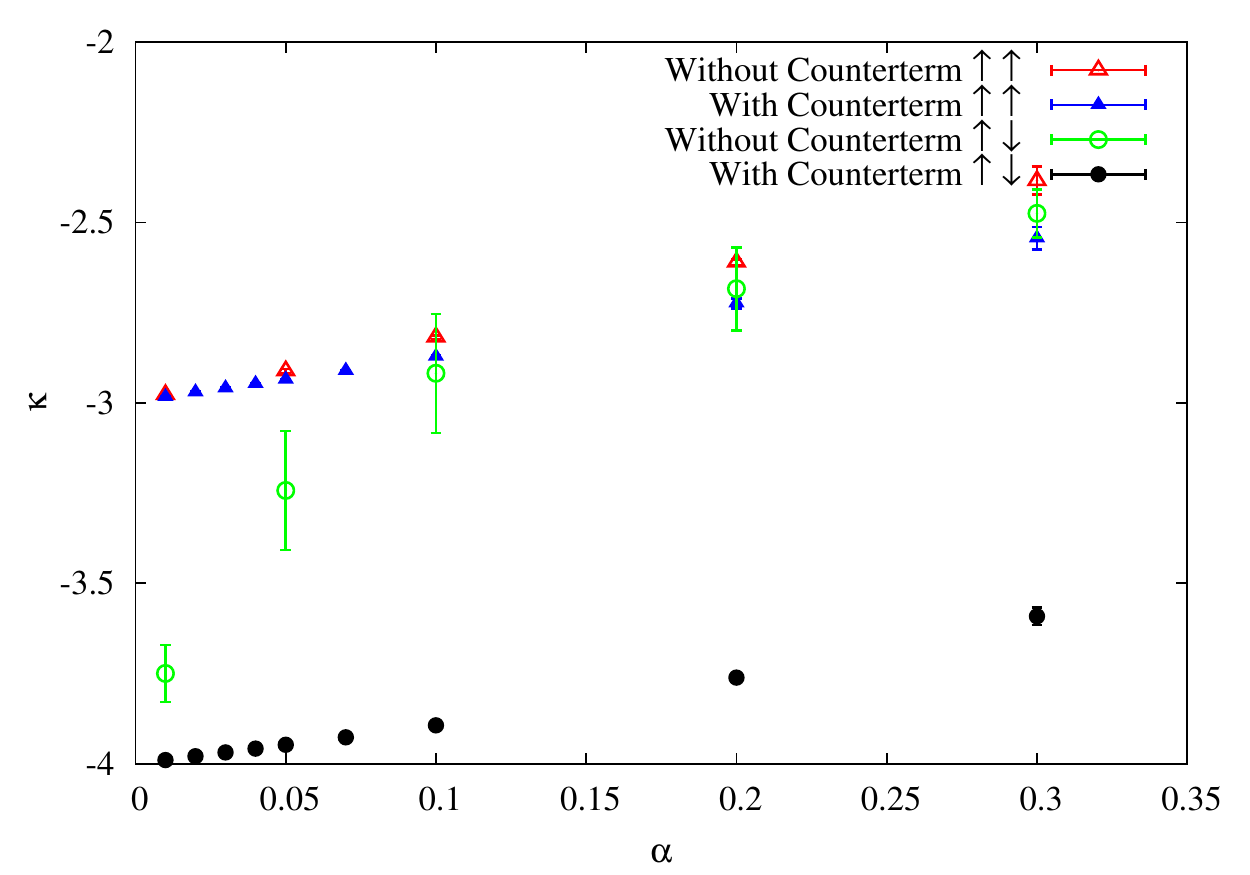}
 \caption{\label{fig:kappaalpha}$\kappa$ vs.\ $\alpha$ for the
 $\uparrow\uparrow$ and $\uparrow\downarrow$ components of the
 $1^1S_0$ state.  Open (closed) symbols indicate results excluding
 (including) the regularization term.}
\end{figure}

Using our entire set of $\alpha$ results, it is possible to study the
effect of varying $\alpha$ upon $\kappa$.  Shown in
Fig.~\ref{fig:kappaalpha} are the extracted values of $\kappa$ for
both the dominant $\uparrow\downarrow$ component of the singlet state
and the subleading $\uparrow\uparrow$ component.  We have also
obtained values of $\kappa$ for a smaller set of $\alpha$ without
using our regularization scheme.
Because the regularization term is only needed for $G_2$, it makes
sense that only the $\uparrow\downarrow$ has a dramatic change in its
$\alpha$ dependence by the introduction of the regularization term,
whereas the $\uparrow\uparrow$ wave functions are mostly unaffected.

\subsection{Decay Constants}
In addition to the invariant masses, the decay constants offer an
interesting observable that can be extracted from the wave functions.
They also serve as an good test bed for understanding how the
properties of the wave function are affected by regularization and
renormalization.  The decay constants in the vector $V$ and
pseudoscalar $P$ channels are defined by
\begin{align}
 \langle0|\bar{\psi}\gamma^\mu\psi|V(p),\lambda\rangle&=
\epsilon^\mu_\lambda m_V f_V,\nonumber\\
 \langle0|\bar{\psi}\gamma^\mu\gamma^5\psi|P(p)\rangle&=ip^\mu f_P, 
\end{align}
where $\epsilon^\mu_\lambda(p)$ is the polarization vector for the
boson, and $\lambda=0,\pm1$.  In front-form field theory, the decay
constants can be computed directly from the $+$ components of these
currents which, following Ref.~\cite{Branz:2010ub,Li:2015zda}, are
given for QED bound states by
\begin{widetext}
\begin{align}
 &f_{V(P)}=\int \frac{\de x}{\sqrt{x(1-x)}}
\frac{\de ^2\bm{k}_\perp}{(2\pi)^3}\left[\psi^J_{J_z=0}
(\bm{k}_\perp,x,\uparrow\downarrow)\mp\psi^J_{J_z=0}
(\bm{k}_\perp,x,\downarrow\uparrow)\right],
\end{align}
\end{widetext}
where the vector (pseudoscalar) decay constant is given by the
difference (sum) of the two terms in the equation.  Taking the
component wave functions from TMSWIFT calculations, it is possible to
obtain $f_V$ for the singlet state and $f_P$ for the triplet state as
a function of $\alpha$.  Like the invariant masses, the decay
constants are found to be well fit to the functional form of
Eq.~(\ref{eq:mfit}), and therefore an infinite-cutoff values for them
can be obtained.  These results can be found in Table~\ref{tab:esp}.

For the decay constants, one expects
$f_i\propto|\psi_i(0)|/\sqrt{M_i}$, which suggests a $\alpha^{3/2}$
power law at leading order.  To check this prediction, a fit is
performed to the function
\begin{equation}
\label{eq:ffit}
 f_i(\alpha)=N\alpha^\beta,
\end{equation}
and the results are exhibited in Table~\ref{tab:falpha}\@.  

\begin{table}[ht]
\begin{center}
  \caption[Fit Parameters of Eq.~(\ref{eq:ffit}) for the Vector Decay
  Constant of the Singlet State and the Pseudoscalar Decay Constant of
  the Triplet State for Two Ranges of $\alpha$.  The Leading-Order
  Perturbative Prediction Is $\beta=3/2$.]{Fit parameters of
    Eq.~(\ref{eq:ffit}) for the vector decay constant $f_V$ of the
    singlet state and the pseudoscalar decay constant $f_P$ of the
    triplet state for two ranges of $\alpha$.  $N$ has units of $m$.
    The leading-order perturbative prediction is $\beta=3/2$.}
\label{tab:falpha}
\begin{tabular}{l|c|c|c}
\hline\hline
$f_i$&$\alpha$&$N$&$\beta$\\ \hline
$f_V$ & [0.01,0.3] & 0.0412(9) &  1.510(7) \\
     & [0.01,0.1] & 0.0411(3) &  1.509(3) \\
$f_P$ & [0.01,0.3] & 0.022(3) &  1.37(4) \\
     & [0.01,0.1] & 0.0240(8) &  1.394(10) \\
\hline
\end{tabular}
\end{center}
\end{table}

Similar to the invariant masses, the $f_V$ values for the singlet
state seem to reproduce the perturbative form to leading order very
well over for all values of $\alpha$.  The agreement between $f_P$ for
the triplet state shows a poorer agreement, especially for large
$\alpha$, where the inclusion of the annihilation channel enables
higher-order corrections to the decay rate.

\subsection{Multiple-Flavor Effects} \label{subsec:Multiple}

True muonium is acutely sensitive to the effects of multiple flavors.
The large mass difference $m_\mu/m_e\approx207$ causes electronic loop
corrections to be the largest corrections to the spectrum of true
muonium.  Additionally, the ratio $m_\tau/m_\mu\approx16$ is small
enough to produce appreciable effects on the system at
$\mathcal{O}(\alpha^5)$.  While the vacuum polarization in the
exchange diagrams is neglected by our model, it is possible to study
these effects in the annihilation channel.

\begin{figure}[h]
\includegraphics[width=\linewidth]{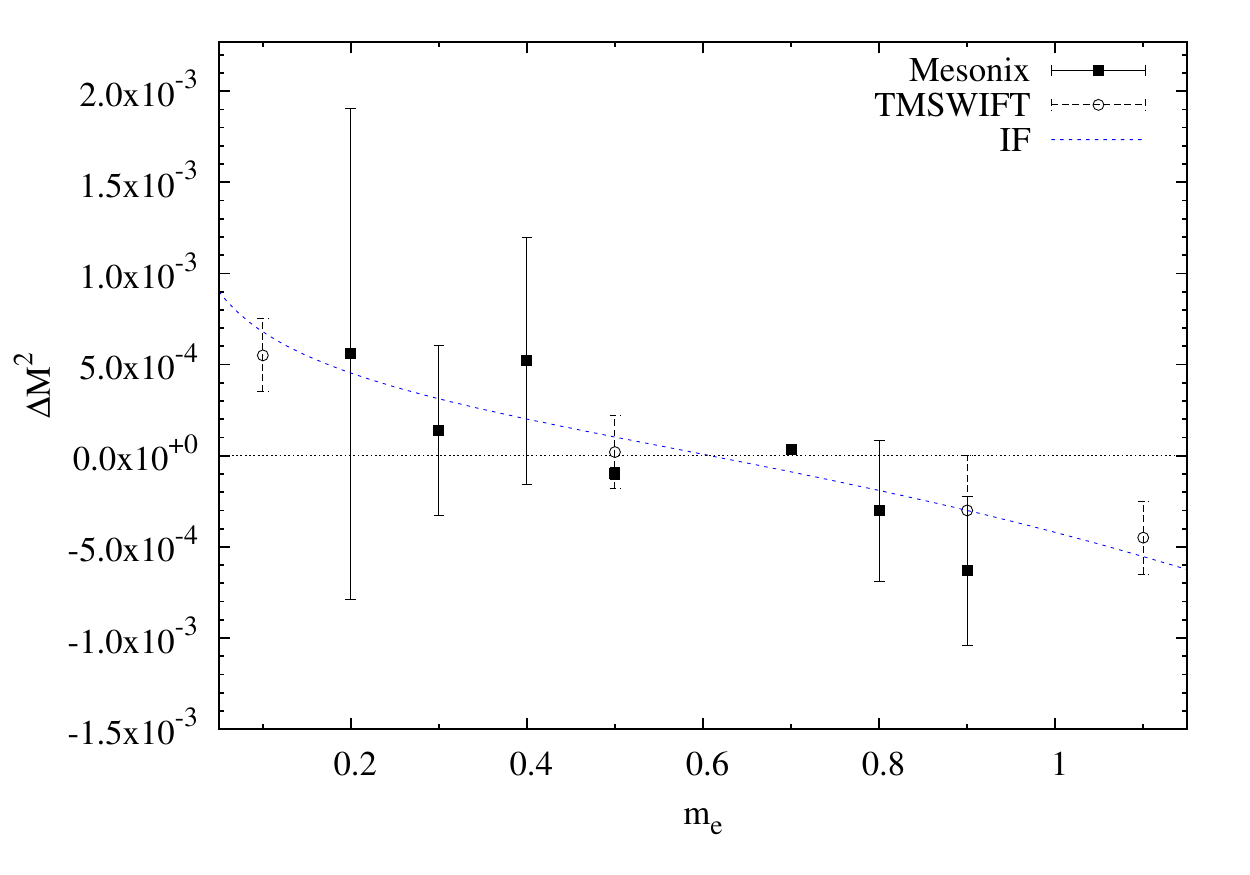}
\caption{$\Delta M^2$ corrections from a second flavor of leptons to
  true muonium as a function of the second flavor's mass $m_e$.
  Errors are estimated from the numerical fit alone.}
\label{fig:eloop}
\end{figure}

\begin{figure*}
\includegraphics[width=\linewidth]{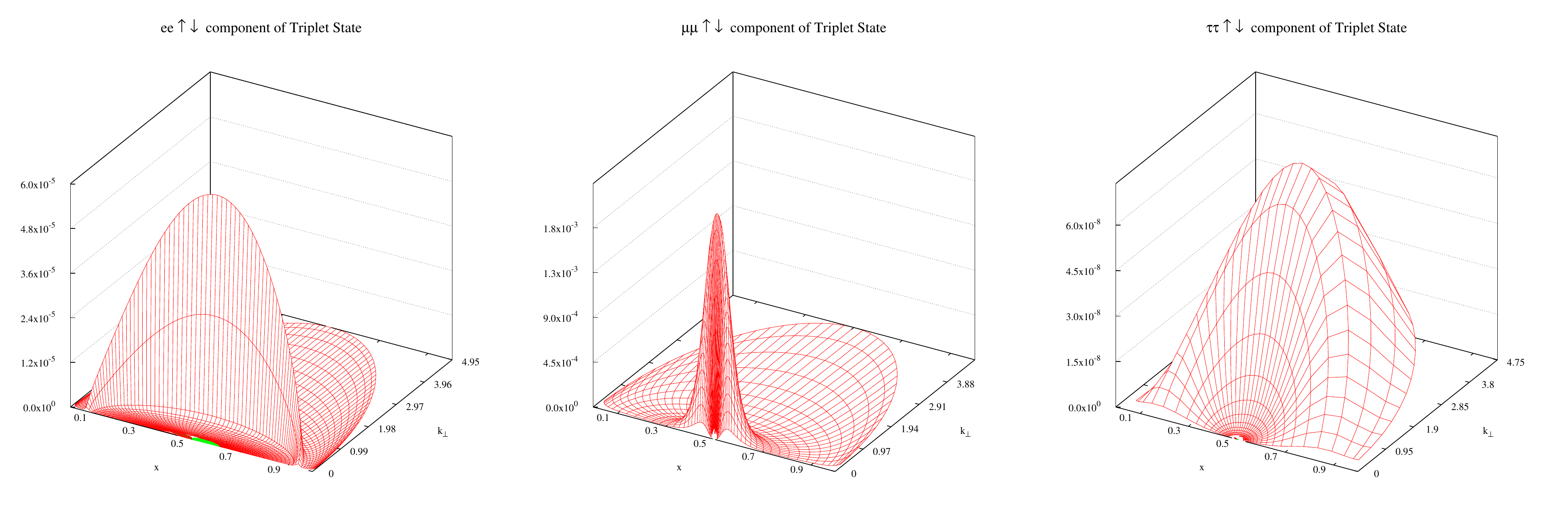}
 \caption{\label{fig:waves}The $1^3\!S_1^0$ probability density of the
 $\uparrow\downarrow$ components of (left) $\ee$, (center) $\mm$, and
 (right) $\tta$, with $J_z=0$, as a function of $x$ and $k_\perp$, for
 $\alpha=0.3$, $m_\mu/m_e=m_\tau/m_\mu=2$, $\Lambda_i=5\alpha m_i$,
 and $N_\mu=37$, $N_\tau=31$, $N_e=71$.}
\end{figure*}
Previous results~\cite{Lamm:2013oga} found large, nonlinear $N$ and
$\Lambda_e$ dependence from the electronic contribution, even for the
unphysically large ratio of $m_\mu/m_e=2$.  With TMSWIFT, we have been
able to further study this dependence.  Numerical limitations prevent
the collection of a sufficiently large number of simulations to fit to
Pad\'{e}-approximants.  Instead, we fix $\alpha = 0.3$, $N_\mu=21,
\Lambda_\mu=10$ ($\Lambda_i$ is given in units of $m_i \alpha$), and
then obtain estimates for $\Delta M^2$ (the shift of squared mass
eigenvalues due to the inclusion of additional lepton flavors) by
averaging over the ranges $N_e \in [27,35]$ and $\Lambda_e \in
[1,35]$.

We have been able to further reduce the uncertainty through two new
ideas.  First, simulations were made using two different
discretization schemes, Gauss-Legendre and Curtis-Clenshaw.  The use
of two discretization schemes for the same $N_e$ allows us to explore
the effects of discretization on the continuum electron states with
smaller $N$.  Additionally, $f_P$ is a sensitive probe of the coupling
of electron continuum states to the bound state.  Empirically, we find
that if the value of $f_P$ differs by more than 10\% from the
single-flavor case, the simulation has sampled the continuum in an
inaccurate way and can be excluded from the average.

Producing results for the physical value of the electron mass remains
difficult numerically because of the large separation of scales.  Our
results for the corrections to true muonium from electronic loops in
the annihilation channel are shown in Fig.~\ref{fig:eloop}, compared
to the anticipated instant-form result, and the previous results
of~\cite{Lamm:2013oga}.  One can see that TMSWIFT's parallel
implementation, while still numerically limited, can produce better
agreement with the instant form than found in~\cite{Lamm:2013oga},
with smaller uncertainty.

TMSWIFT has also been written to allow for an arbitrary number of
flavors.  We present here results from a three-flavor true muonium
model, albeit with unphysical ratios $m_\mu/m_e=m_\tau/m_\mu=2$,
keeping $\alpha=0.3$.  In Fig.~\ref{fig:waves} are shown the
probability densities of the $\uparrow\downarrow$ components of each
flavor for the triplet state.  In Table~\ref{tab:prob} we present the
relative probability for each component in this case.
\begin{table}[ht]
\caption{Integrated probability for each flavor in the true muonium
  $1^3S_1^0$ state.  The parameters used are $\alpha=0.3$,
  $m_\mu/m_e=m_\tau/m_\mu=2$, $\Lambda_i=5\alpha m_i$, and
  $N_\mu=37,N_\tau=31,N_e=71$.}
\begin{tabular}{l|c}
\hline\hline
Flavor & $\int\mathrm{d}x \, \mathrm{d}^2 \! \bm{k}_\perp
P(x,{k}_\perp)$\\ \hline
$|\mm\rangle$&$0.992$\\
$|\ee\rangle$&$0.008$\\
$|\tta\rangle$&$\approx 1.2\times10^{-5}$\\
\hline
\end{tabular}
\label{tab:prob}
\end{table}

\section{Discussion and Conclusion}
\label{sec:cons}
In this work we have presented results for the invariant mass and
decay constants of the true muonium system.  For the first time, we
have gone beyond the case $\alpha=0.3$ and shown that the approach to
$\alpha_{\rm QED}$ is possible with sufficient numerical resources.
The purpose of this program is not to produce energy levels
competitive in the weak-field limit with perturbative calculations.
Instead, our goals in calculating at $\alpha_{\rm QED}$ are to produce
true muonium wave functions that can be used in relativistic
situations, and as to provide an independent check on our methods,
allowing one to be confident in the strong-field predictions.
Furthermore, using our previously developed regularization scheme, the
simultaneous limits of $N\rightarrow\infty$ and
$\Lambda\rightarrow\infty$ have been taken and stable results found.
These values have been compared to the instant-form perturbative
calculations, and reasonable agreement has been obtained.  Finally,
initial studies have been undertaken to compute the fully
nonperturbative contribution to the bound state arising from
additional flavors, both lighter and more massive then the muon.
Improved agreement with instant-form predictions have been obtained
for a range of masses of a second flavor, and simulations of the
three-flavor model have been produced.

Currently, work is underway to include the $|\gamma\gamma\rangle$
state and the pair of states $|\ell\bar{\ell}\ell\bar{\ell}\rangle$
and $|\ell\bar{\ell}\ell'\bar{\ell}'\rangle$, which are required for
gauge invariance.  These corrections are crucial for precision true
muonium predictions and are a necessary step for QCD bound states as
well.

Proper renormalization of the Hamiltonian is the remaining obstacle.
In order to make accurate predictions, the $\Lambda$ dependence found
in this work must be systematically removed, which involves not just
including new Fock sectors, but imposing gauge invariance at each
stage.  A proper implementation of charge renormalization and the
running of the coupling $\alpha$ should address a large part of the
issue.  A first step in this direction would focus upon implementing a
renormalized vacuum polarization into the effective interactions.
With a robust renormalization scheme, multiple values of $\Lambda$
would not be needed to take the $\Lambda\rightarrow\infty$, greatly
reducing the numerical effort to produce reliable results.  With
TMSWIFT, Fock-space limitations have been greatly decreased.  This
improvement allows for the implementation of explicit Fock-state
renormalization methods like Pauli-Villars
regulators~\cite{Chabysheva:2009vm,Chabysheva:2010vk,
  Malyshev:2013eca} and sector-dependent
counterterms~\cite{Karmanov:2008br, Karmanov:2012aj}.  Using the
exchange properties of leptons could further reduce the number of
basis states, similar to the methods used in
Ref.~\cite{Chabysheva:2014rra} for bosons.  More time-intensive
renormalization schemes like the Hamiltonian-flow
method~\cite{Gubankova:1998wj,Gubankova:1999cx} also become viable
with a parallel implementation.

\begin{acknowledgments}
This work was supported by the National Science Foundation under
Grants PHY-1068286 and PHY-1403891.  This work used the Extreme
Science and Engineering Discovery Environment (XSEDE), which is
supported by National Science Foundation Grant Number ACI-1053575, and
the ASU Physics Computing Cluster.
\end{acknowledgments}

\bibliography{TM_2_0718}
\end{document}